\documentclass[conference]{IEEEtran}
\pdfoutput=1
\usepackage{cite}
\usepackage{amsmath,amssymb,amsfonts}
\usepackage{algorithmic}
\usepackage{graphicx}
\usepackage{textcomp}
\usepackage{xcolor}
\usepackage[T1]{fontenc}

\begin{document}

\title{Exponential Shift: Humans Adapt to AI Economies}

\author{\IEEEauthorblockN{Kevin J McNamara}
\IEEEauthorblockA{New Jersey, USA \\
Email: kevin@mcnamara-group.com}
\and
\IEEEauthorblockN{Rhea Pritham Marpu}
\IEEEauthorblockA{New Jersey, USA \\
Email: rm422@njit.edu}}

\maketitle

\begin{abstract}
This paper explores how artificial intelligence (AI) and robotics are transforming the global labor market. Human workers, limited to a 33\% duty cycle due to rest and holidays, cost \$14 to \$55 per hour. In contrast, digital labor operates nearly 24/7 at just \$0.10 to \$0.50 per hour. We examine sectors like healthcare, education, manufacturing, and retail, finding that 40-70\% of tasks could be automated. Yet, human skills like emotional intelligence and adaptability remain essential. Using real-world examples, this paper proposes strategies—like a 4-day workweek and retraining—to ensure a fair transition to an AI-driven economy.
\end{abstract}


\section{Introduction}
The rapid rise of artificial intelligence (AI) and robotics is fundamentally changing the way we work. Humans are constrained by the need for rest, sleep, and time off, limiting their productive capacity to about one-third of the day (a 33\% duty cycle). Digital systems, on the other hand, can operate almost continuously, offering significant cost savings. However, this shift raises concerns: jobs may be displaced, energy consumption may surge, and societies must adapt to new economic realities. Central to this transformation is the increasing automation of information processing tasks across industries. Many professional jobs involve generating, analyzing, and transforming information in ways that AI systems can now assist with or perform independently. This paper compares human and digital labor, examines how jobs will change, and explores sector-specific impacts. It also addresses the energy and ethical challenges of automation and proposes strategies for a fair transition.

\section{What are Tokens and Why Do They Matter?}
In AI, tokens are the building blocks of text-words, parts of words, or characters-that models process to generate or analyze language. For example, the sentence "AI is transforming jobs" may break into tokens such as "AI," " is," " transform," "ing," and " jobs," depending on the tokenizer and the specific model being used. Humans can process about 5,000 to 20,000 tokens per hour, limited by cognitive and physical constraints [5], [6]. AI, however, can process millions of tokens at the same time, making it highly efficient for tasks like drafting documents or analyzing data. But AI cannot replicate human creativity or ethical reasoning, meaning it complements rather than replaces many roles. Understanding tokens helps us see where AI can boost productivity and where human skills remain irreplaceable.

\section{Comparing Human and Digital Labor}
Consider the fundamental differences between human and digital labor: where human workers require legal protections for rest and recovery, AI systems demand only minimal maintenance to sustain performance. As businesses and industries increasingly leverage this digital workforce, productivity ceilings once dictated by human limitations are being redefined, potentially ushering in an era of unprecedented economic potential. This comparison sets the stage for understanding AI's exponential growth--not merely as a tool, but as a foundational force redefining the very nature of labor in our modern economy. 

\subsection{Key Differences}
The contrast between human need for recovery and AI's minimal maintenance requirements is reshaping productivity limits across industries.
\begin{itemize}
    \item \textbf{Duty Cycle:} 
    The fundamental rhythm of work differs dramatically between humans and machines, creating an efficiency gap that transforms business operations.
    \begin{itemize}
        \item \textbf{Humans:} Work at 33\% capacity due to 8-hour days and time off.[1][7]
        \item \textbf{Digital Labor:} Operates at nearly 100\%, with only brief maintenance periods and occasional updates.[8]
    \end{itemize}
    \item \textbf{Breaks and Holidays:} 
    Calendar constraints create a stark productivity divide, as the biological necessities of human workers contrast with the relentless availability of digital systems.
    \begin{itemize}
        \item \textbf{Humans:} Need regular rest and get 10-15\% fewer working days due to holidays and leave.[7]
        \item \textbf{Digital Labor:} Can function continuously, including weekends and holidays.
    \end{itemize}
    \item \textbf{Cost:} 
    The economics of labor reveal perhaps the most compelling case for automation, with a price differential that continues to widen as technology advances.
    \begin{itemize}
        \item \textbf{Humans:} \$14-\$55/hour, depending on the job, skill level, and geographical location.
        \item \textbf{Digital Labor:}  \$0.10-\$0.50/hour for basic text processing, with costs varying based on model complexity, hardware requirements, and usage volume. These costs continue to decline as technology improves.
    \end{itemize}
\end{itemize}

\begin{figure}[h!]
    \centering
    \includegraphics[width=\columnwidth]{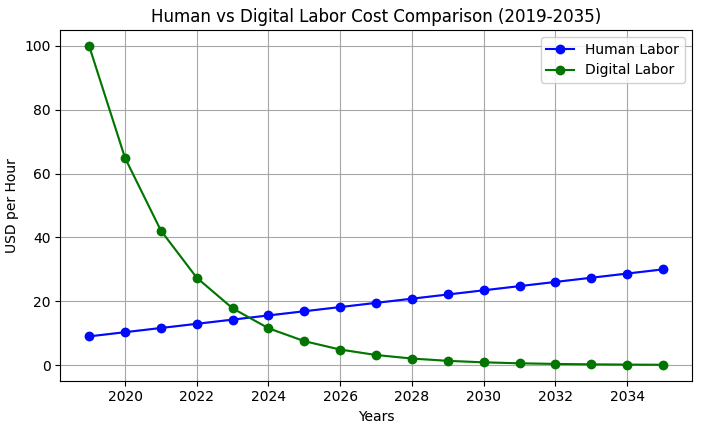}
    \caption{This line graph illustrates the dramatic cost inversion between human and digital labor over a 16-year period, with human wages steadily rising with inflation while digital labor costs plummet as hardware efficiency improves and computational power becomes increasingly affordable.}\label{fig 1:cost_comparison}
\end{figure}

\begin{figure}[h!]
    \centering
    \includegraphics[width=\columnwidth]{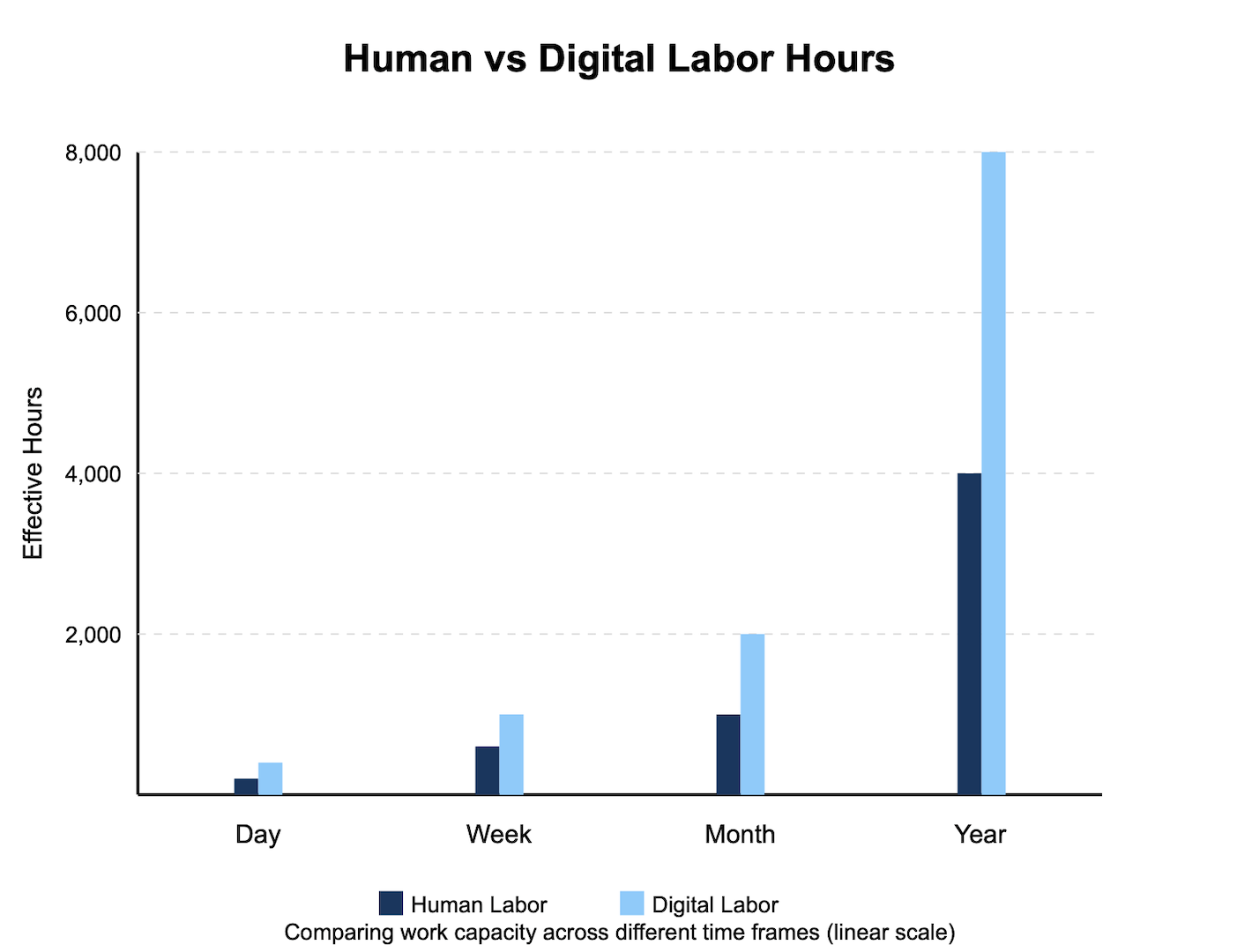}
    \caption{This bar chart illustrates the striking disparity between human and digital labor capacity across time frames, with digital systems operating almost continuously while human workers remain limited by biological necessities and legal protections.}\label{fig 2:your_graph}
\end{figure}

\subsection{Token Usage in Job Roles}
To understand how AI impacts various professions, it’s useful to measure the text-based workloads that dominate many roles—workloads that can be quantified in tokens, the basic units of text processed by AI systems. Table 1 below estimates the number of tokens processed per hour across a range of job roles, reflecting tasks such as writing, coding, or analyzing data. These figures are not exact measurements but approximations inspired by productivity research and text-generation patterns. They draw from studies like MIT’s "The Work of the Future" [17], which examines job task automation, and Stanford’s "Artificial Intelligence Index Report 2024" [18], which explores AI’s text-processing capabilities. Additional insights come from Microsoft Research’s "AI and Productivity in the Workplace" [21] and OpenAI’s "Token Usage Patterns in AI Engineering and Research" [20], which provide data on human and AI text-handling capacities. By comparing these token estimates, we can see where AI might augment or replace human effort, offering a lens into the evolving division of labor.

\begin{table}[h!]
    \caption{Estimated Tokens (units of text processed by AI) and Tasks by Job Role}
    \centering
    \begin{tabular}{|p{2cm}|c|p{4cm}|}
    \hline
    \textbf{Job Role} & \textbf{Tokens / Hour} & \textbf{Tasks Involving Tokens} \\ \hline
    Administrative Assistant & 1,315 & Scheduling, email drafting, document updates \\ \hline
    AI Engineer & 2,667 & Designing, training, and maintaining AI models \\ \hline
    AI Researcher & 3,000 & Experiments, academic papers, literature reviews \\ \hline
    Business Analyst & 1,567 & Report writing, data analysis, presentations \\ \hline
    Cloud Architect & 1,834 & Cloud infrastructure design, documentation \\ \hline
    Content Writer & 2,500 & Article writing, content creation \\ \hline
    Customer Service Representative & 1,200 & Email and chat support, ticket responses \\ \hline
    Data Analyst & 2,000 & Data reports, query writing, visualizations \\ \hline
    Financial Analyst & 2,500 & Financial reports, data analysis, forecasts \\ \hline
    HR Manager & 1,800 & Job descriptions, performance reviews, emails \\ \hline
    Journalist & 2,000 & Article writing, research summaries \\ \hline
    Legal Assistant & 1,500 & Document review, legal drafting \\ \hline
    Marketing Specialist & 2,000 & Ad copy, social media posts, marketing plans \\ \hline
    Medical Professional & 2,000 & Patient notes, reports, research papers \\ \hline
    Software Developer & 3,000 & Coding, debugging, documentation \\ \hline
    Teacher & 1,500 & Lesson plans, feedback, educational content \\ \hline
    \end{tabular}
    \footnotesize
    \vspace{1mm}\label{tab:job_roles}

\small{Note: These estimates are approximate and based on typical text-generation tasks for each role. For context, modern AI systems can process millions of tokens per hour, representing a 500-1000x increase over human processing rates.}
\end{table}

\subsection{Strengths and Weaknesses}
The comparison between human and digital labor hinges on more than just raw output--it's about their distinct capabilities and limitations in real-world applications. Table 2 breaks down these differences across key dimensions like efficiency, cost, and sustainability, revealing a clear divide: digital labor thrives in relentless, repetitive tasks, whilehumans excel in nuanced, creative, and empathetic roles. This framework builds on research from MIT's "The Work of the Future" [17], which highlights human adaptability in complex problem-solving, and Stanford's "Artificial Intelligence Index Report 2024" [18], which notes AI's superiority in high-volume data processing. 

Case studies illustrate these dynamics in action: Amazon's warehouse robots cut costs but require constant power, while the Associated Press's AI-generated earnings reports free journalists for deeper storytelling. Studies like Strubell et al. [4] further underscore AI's energy demands, contrasting with human sustainability limits noted in the ILO's work-life balance report [7]. This analysis shows how each form of labor complements the other, shaping a hybrid workforce where strengths are maximized and weaknesses mitigated.

\begin{table}[h!]
    \caption{Comparison of Human and Digital Labor}
    \centering
    \begin{tabular}{|p{2cm}|p{2.5cm}|p{2.5cm}|}
    \hline
    \textbf{Aspect} & \textbf{Human Labor} & \textbf{Digital Labor} \\ \hline
    Efficiency & Good at creative and complex tasks & Great for repetitive, high-volume work \\ \hline
    Cost & Includes wages and benefits & Needs upfront investment, no benefits \\ \hline
    Sustainability & Needs rest to stay productive & Runs endlessly with power and upkeep \\ \hline
    Limits & Gets tired, works limited hours & Can’t handle empathy or moral choices \\ \hline
    Best Use & Problem-solving, caring roles & Data crunching, assembly lines \\ \hline
    \end{tabular}
    \footnotesize
    \vspace{1mm}\label{tab:labor_comparison}
    \small{Note: Comparative Analysis of Human and Digital Labor Capabilities Across Key Performance Dimensions.}

\end{table}

\section{How Jobs Will Change}
Imagine a world where the hum of machinery replaces the chatter of workers on a construction site, where trucks roll across highways guided by algorithms instead of human hands, and where office cubicles grow quieter as AI silently takes over routine tasks. This isn’t a distant sci-fi vision—it’s the near future, unfolding as artificial intelligence and robotics redefine the labor market. The shift is seismic: repetitive jobs face obsolescence, white-collar roles bend under the pressure of automation, and entirely new professions emerge from the technological frontier. Drawing from McKinsey’s [2], Medium “The future of work in the age of AI” [13] and ARK Invest’s "Big Ideas 2024" [3], this section explores the jobs most at risk, those in transition, and the opportunities arising amid the upheaval. It’s a story of disruption and adaptation, where human ingenuity must find its place alongside relentless digital efficiency.

\subsection{Blue-Collar Jobs at Risk}
Machines don’t tire, don’t unionize, and don’t need coffee breaks—qualities that make them formidable competitors for blue-collar workers in repetitive, predictable roles. The data paints a stark picture:
\begin{itemize}
    \item \textbf{Construction Workers (\$26.93/hour):}  Hammering nails and stacking bricks could soon be augmented by technology, with McKinsey [15] estimating that 70\% of assembly tasks could be automated by 2035 [2]. Robotic arms, like those tested by companies such as Boston Dynamics, already lay bricks with precision, potentially transforming millions of jobs in the trades.
    \item \textbf{Truck Drivers (\$22.00/hour):} The open road, once a symbol of freedom, may belong to self-driving fleets by 2030. ARK Invest predicts autonomous trucks will dominate long-haul routes, slashing costs and eliminating human error [3]. Companies like Tesla and Waymo are racing to deploy these vehicles, putting 3.5 million U.S. trucking jobs at risk.
    \item \textbf{Retail Cashiers (\$14.00/hour):} Self-checkout kiosks and AI-driven inventory systems are already shrinking cashier lines. McKinsey [15] forecasts that half of these roles could vanish by 2030 as retailers like Walmart and Amazon Go pioneer cashierless stores [2].
    \item  Yet, not all blue-collar work is doomed. Jobs demanding flexibility and on-the-spot problem-solving—like plumbers diagnosing a leak or electricians rewiring a faulty circuit—will endure. These trades will increasingly wield AI tools, such as diagnostic apps or augmented reality guides, but the human touch remains irreplaceable where adaptability trumps repetition.
\end{itemize}

\subsection{White-Collar Jobs in Flux}
The office isn’t immune to AI’s reach. As algorithms master data crunching and text generation, white-collar roles built on routine processes face a reckoning. Here’s where the changes are emerging:
\begin{itemize}
    \item \textbf{Administrative Assistants (\$20.00/hour):}  Scheduling meetings, drafting emails, and updating spreadsheets—tasks that once filled an assistant’s day—are now fair game for AI. McKinsey suggests 40\% of these duties could be automated by 2030 [2]. Tools like Microsoft’s Copilot already handle such chores, shrinking the need for human oversight.
    \item \textbf{Accountants (\$35.00/hour):}  Bookkeeping, tax prep, and basic audits are being streamlined by software like QuickBooks and AI platforms from Deloitte. McKinsey indicates that routine accounting tasks face high automation potential, though high-level analysis and client advising will keep accountants relevant [2].
    \item \textbf{Customer Service Representatives (\$15.00/hour):} “How may I assist you today?” might increasingly come from a chatbot. With AI handling 80\% of routine calls by 2030, per McKinsey [2] and U.S. Bureau of Labor Statistics [14], firms like Zendesk are deploying bots that resolve complaints faster than humans can pick up the phone. Still, escalations requiring empathy or complex judgment will preserve some human roles.
    \item  Not all white-collar jobs will fade—strategic positions like managers, who navigate ambiguity and inspire teams, are poised to adapt rather than disappear. These roles will lean on AI for insights, not replacements, evolving into hybrid functions where human intuition guides machine precision.
\end{itemize}

\subsection{New Opportunities}
Amid the disruption, AI isn’t just a job-killer—it’s a job-creator, birthing roles that didn’t exist a decade ago. As industries harness digital labor, new frontiers emerge:
\begin{itemize}
    \item \textbf{AI Specialists (\$60.00/hour):} The architects of this revolution are in high demand. McKinsey projects a 15\% growth in AI-related jobs by 2030 [2], from engineers training models to ethicists ensuring fairness. Companies like Google and OpenAI are snapping up talent to push AI’s boundaries.
    \item \textbf{Data Scientists (\$50.00/hour):} With AI churning out oceans of data, those who can sift through it for meaning are gold. Demand is expected to surge 40\% by 2030 [14][15][2], as firms from healthcare to finance—like IBM and Goldman Sachs—rely on data scientists to turn raw numbers into strategy.
    \item \textbf{Energy Experts (\$42.00/hour): } AI’s voracious power appetite, detailed in Section VI, spawns a need for sustainability wizards. These professionals will design green solutions—like small nuclear reactors or solar grids—to offset the 3.5 to 7 times higher energy costs of digital labor [4]. Roles at startups like Commonwealth Fusion Systems exemplify this growing niche.
    \item This wave of opportunity demands new skills: coding, data literacy, and energy management become the currency of the future. Governments and businesses must act swiftly to retrain workers, ensuring the displaced can ride this rising tide rather than be swept away.
\end{itemize}

\section{Sector-Specific Impacts and Example Use Cases}
Step into a future where a doctor’s diagnosis is sharpened by an algorithm [15], a student’s lesson adjusts to their curiosity in real time [12], and a store checkout hums along without a cashier in sight [15]. This isn’t a speculativedream—it’s the reality AI is forging across healthcare, education, manufacturing, and retail, transforming how we heal, learn, build, and buy. The numbers tell a compelling story: cost savings soar and productivity leaps [14], [15],yet the human element persists where machines falter [15]. Anchored in McKinsey’s analysis of automation impact [15], UNESCO’s education tech insights [12], and Strubell et al. 's energy cost findings [5], this section peels back the layers of AI’s sectoral reach. From Amazon’s tireless warehouse bots to the Associated Press’sAI-powered newsroom [15], real-world cases spotlight the technology’s reach—and its limits—offering a roadmap to an economy where digital and human labor must coexist.

\begin{figure}[h!]
    \centering
    \includegraphics[width=\columnwidth]{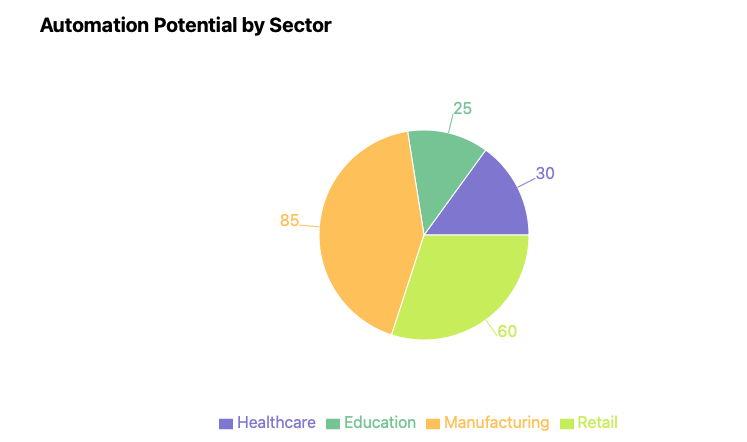}
    \caption{Manufacturing dominates automation potential at 85\%, while education remains most resistant at just 25\%, with retail and healthcare at intermediate levels.}
    \label{fig 3:your_graph}
\end{figure}

\subsection{Healthcare}
AI is revolutionizing healthcare with precision, tackling paperwork and supporting diagnostics to deliver cost savings of up to 30\%, according to McKinsey [2]. Picture a bustling hospital: AI scheduling tools, like those from Epic Systems, optimize operating room slots, sidelining clerks who once juggled calendars by hand [2]. Diagnostic software, such as Google Health’s AI that detects breast cancer in mammograms with accuracy rivaling radiologists, empowers doctors to focus on patient care rather than data drudgery [2]. Yet, the operating theater and bedside remain human domains—surgeons wielding skill honed by years of practice and nurses offering compassion no code can mimic stay irreplaceable [2].
\newline\textbf{Energy Note:}This efficiency comes with energy considerations. Hospital AI systems, from servers to imaging tools, could significantly increase energy use, potentially offsetting savings, per Strubell et al. [4]. Balancing this trade-off will define healthcare’s AI future.

\subsection{Education}
In classrooms and online platforms, AI is rewriting the textbook on learning, delivering personalized lessons and administrative relief that could save 20-25\% per student, as UNESCO reports [12]. Platforms like Coursera harness AI to generate quizzes tailored to a student’s progress, freeing teachers from rote grading to mentor and inspire [12]. Imagine a middle school where an AI tutor adjusts math problems to a child’s pace, or a university where admin bots handle enrollment, letting professors dive deeper into discussion. Still, the human spark—guiding a struggling student or igniting a love for literature—remains beyond AI’s grasp [12].
\newline\textbf{Energy Note:} The shift to digital learning isn’t power-neutral; online systems could significantly increase school energy consumption, a hidden cost as education scales into the cloud [4].

\subsection{Manufacturing}
The factory floor is AI's playground, where robots obliterate labor costs for repetitive tasks by 80-90\%, per McKinsey [2]. Amazon's warehouse bots, rolling 24/7 through fulfillment centers, stack boxes and pick orders with relentless precision, turning what once took shifts of workers into a seamless dance of steel and code [2]. This isn't just theory--Kiva robots have slashed Amazon's operating costs, proving the model's might [2]. But the catch lies in the power grid: a fleet of 100 robots guzzles energy, adding significant costs annually to the bill [4]. Human workers, meanwhile, retain their edge in roles like quality control, where intuition spots flaws machines miss [2]. Manufacturing's future hinges on marrying robotic stamina with human oversight.  
\newline\textbf{Energy Note:} While manufacturing automation delivers impressive efficiency, its environmental footprint extends beyond electricity consumption to include cooling systems, maintenance infrastructure, and backup power requirements. Studies indicate these hidden energy costs can impact regional power grids, making sustainable energy planning a strategic priority for manufacturers in energy-constrained markets [5].

\subsection{Retail}
Retail is shedding its human skin for AI-driven efficiency, with self-checkouts and stock bots trimming costs by 60\% [2]. Walk into an Amazon Go store: cameras and sensors track your picks, charging you as you leave—no cashier required [2]. Stock bots, like those from Walmart, glide through aisles, restocking shelves faster than any clerk [2]. Yet, the human touch endures where trust is currency—a salesperson’s smile or a manager’s empathy during a return keeps customers loyal in ways algorithms can’t replicate [2].
\newline\textbf{Energy Note:} This convenience isn’t free; retail AI could significantly lift energy bills, a price tag that tempers the bottom-line gains [4].

\subsection{Real-World Use Cases}
AI’s fingerprints are already on the world, reshaping specific fields with striking results:  
\begin{itemize}
    \item \textbf{Journalism:} Tools like GPT models churn out data-driven stories at scale. The Associated Press uses AI to produce 3,700 earnings reports each quarter, a task that once bogged down reporters. Now, journalists chase investigative scoops while AI handles the numbers. Limitation: Nuanced storytelling—capturing a source’s emotion or uncovering hidden angles—eludes AI, leaving editors to polish the prose.  
    \item \textbf{Law:} Platforms like ROSS Intelligence draft contracts and briefs in minutes, sifting through legal texts faster than any paralegal. But case law interpretation and client counseling demand human judgment—lawyers remain the ethical compass AI lacks. Limitation: Without moral reasoning, AI’s drafts are tools, not decisions.
    \item \textbf{Software Development:} GitHub Copilot accelerates coding by 30\%, suggesting snippets that turn ideas into lines of code [2]. Developers at firms like Microsoft lean on it for speed, but complex architecture and debugging still call for human ingenuity [2]. Limitation: AI’s suggestions often need tweaking, proving it’s a partner, not a replacement [2].
\end{itemize}

\section{The Energy Downside of Digital Labor}
Imagine a world powered by AI, where every click, calculation, and robotic arm hums with efficiency—yet beneath this sleek surface lies a ravenous energy beast. Digital labor’s promise of cost savings is real, but its power consumption threatens to erode those gains, presenting a critical challenge for humanity. Training a single AI model can produce carbon emissions equivalent to 125 flights from New York to San Francisco, a staggering footprint that dwarfs the modest needs of human labor. Strubell et al. estimate that a 100-unit digital team—such as a fleet of warehouse robots or server racks—consumes \$683,000 to \$1.37 million in energy annually, compared to just \$196,000 for a human crew, representing a 3.5 to 7-fold increase. This energy consumption could diminish the touted 20-40\% cost savings of automation, compelling industries to reassess their financial models.
The implications are significant: if left unchecked, AI’s energy demands could hinder economic progress, disproportionately impacting poorer regions with higher energy costs or outages. However, this challenge presents an opportunity for innovation. Solutions like small modular nuclear reactors and expansive solar arrays could reduce energy costs by 60

\section{Ethical and Social Concerns}
Automation’s march forward isn’t just a technological tale—it’s a human one, fraught with ethical dilemmas and social challenges. Picture a factory worker sidelined by a robot, a rural clerk left behind without tech skills, or an office monitored by AI’s unblinking eye. McKinsey warns that 300 million jobs—nearly one in ten globally—could vanish by 2030 as AI and robotics take hold. This isn’t merely about job loss; it’s a potential widening of wealth gaps, where the tech-savvy thrive while the untrained falter.
Access gaps loom large: workers without retraining, often in underserved areas, risk being locked out of the AI economy, amplifying inequality. Privacy concerns also arise, as AI systems monitoring productivity—like Amazon’s warehouse trackers—could erode personal boundaries, turning workplaces into surveillance hubs.
The impact on society is profound: without intervention, automation could fracture communities, leaving millions economically stranded and distrustful of progress. However, this is a pivotal moment—humans can steer the course. Solutions such as income support, including temporary wage subsidies, and robust skill programs modeled on Germany’s vocational retraining success can cushion the blow. The goal is empowerment: equipping workers to transition into roles that AI cannot easily replicate—creative, caring, or strategic—while ensuring no one is left behind. This isn’t just mitigation; it’s a blueprint for a more equitable digital age.
\begin{itemize}
    \item \textbf{Job Loss:} 300 million jobs at risk by 2030. [2]
    \item \textbf{Access Gaps:} Workers without training could fall behind.
    \item \textbf{Privacy:} AI might track workers too closely, eroding personal boundaries.
    \item \textbf{Potential Fixes:} Income support and skill programs can soften the blow.
\end{itemize}

\section{Strategies for a Fair Transition}
Here are six ways to guide humanity through this shift:
\begin{itemize}
    \item \textbf{4-Day Workweek:} Iceland's large-scale trials between 2015-2019 demonstrated that reduced working hours decreased stress while maintaining or improving productivity.[22]
    \newline Impact: Workers gain time to retrain or rest, preserving mental health as AI takes repetitive loads.    
    \item \textbf{Retraining:} Teaching AI-adjacent skills-critical thinking, coding, ethics-requires focused government initiatives. Programs like Singapore's SkillsFuture and Germany's National Skills Strategy offer current models.[23][24]
    \newline Impact: Displaced workers become AI collaborators, not competitors, securing roles in a tech-driven world.    
    \item \textbf{Universal Basic Income (UBI):} Finland's 2017-2018 basic income experiment showed improvements in recipients' wellbeing and modest employment effects. Similarly, smaller-scale UBI pilots in places like Stockton, California demonstrated how direct cash transfers can provide stability during economic transitions.[12]
    \newline Impact: Financial stability frees humans to pursue education or entrepreneurship, softening automation's blow.     
    \item \textbf{Human Quotas:} Mandating 30\% human staff in automated sectors, as proposed by labor economists, ensures human oversight and employment [10]. 
    \newline Impact: Jobs persist where empathy and judgment reign, balancing efficiency with humanity.    
    \item \textbf{On-Shoring:} Tax incentives for domestic AI development, like those in South Korea's Digital New Deal, help retain technology jobs [12]. Countries could develop layered approaches: subsidies for firms hiring local talent, incentives for domestic AI research, and grants for startups prioritizing regional development. For example, Canada's Global Skills Strategy combines expedited work permits for foreign tech talent with requirements for companies to create domestic employment and training opportunities [13]. 
    \newline Impact: Communities retain economic vitality, countering offshoring's drain, while citizens gain priority access to new roles.    
    \item \textbf{Oversight:} The EU's AI Act establishes a framework for AI risk assessment and ongoing monitoring [14], while the UK's proposed regulatory approach emphasizes principles-based oversight [15]. 
    \newline Impact: Transparent governance builds trust, aligning technological development with societal welfare and ensuring the benefits of AI are broadly shared.    
\end{itemize}

\section{Conclusion}
The data is clear: AI excels in cognitive capabilities—the Intelligence (IQ) dimension—processing information at rates that outpace humans by orders of magnitude while operating at a fraction of the cost. This efficiency drives the economic imperative for automation across sectors from manufacturing to knowledge work. Yet the data reveals that true intelligence operates in multiple dimensions, not all of which yields to algorithmic approaches.

As AI systems edge toward Emotional Quotient (EQ), they remain fundamentally limited in their capacity for genuine empathy. They can mimic emotional responses but cannot truly experience the compassion that guides a healthcare provider or the intuitive understanding that resolves complex workplace dynamics. This gap explains why roles requiring emotional depth remain human-centered despite AI's cognitive advantages.

Moreover, AI is transforming our economy by engaging with both the physical and digital worlds through the "Action Quotient" (AQ)—the capacity to address novel situations without predefined rules. While AI seamlessly automates structured tasks across physical systems and digital frameworks—like APIs, SaaS platforms, bespoke software, and governmental systems—it struggles with the unpredictability that defines both tangible environments and complex digital ecosystems. This shift highlights AI’s economic impact: it optimizes routine processes, yet leaves room for human adaptability to shine in roles requiring real-time problem-solving, whether navigating physical challenges or innovating within dynamic digital landscapes, driving value across industries.

These advanced capabilities come at substantial cost. AI's energy demands—often 3.5 to 7 times higher than human biological processes—threaten to undermine its economic promise, particularly in regions with unstable power infrastructure. This energy premium represents not just an operational challenge but a strategic consideration in determining where and how to deploy these technologies.

The path forward lies not in competition but in complementarity. By pairing human adaptability with machine efficiency, we create systems that transcend the limitations of each. The strategies we've outlined—from restructured work arrangements to targeted skill development—provide a framework for this integration.

The future belongs not to artificial intelligence alone, nor to human capabilities in isolation, but to the thoughtful integration of both. As we advance, the measure of our success will not be how effectively we've automated human labor, but how wisely we've enhanced human potential through technology. This is the true promise of the AI revolution—not replacing humanity, but elevating it.

\end{document}